\documentclass[12pt]{iopart}
\usepackage[dvips]{graphicx}	% required for `\includegraphics' (yatex added)
\usepackage{url}	% required for `\url' (yatex added)
\begin{document}
\title{A sensitive cloud chamber without radioactive sources}
\author{Syoji Zeze$^*$, Akio Itoh, Ayu Oyama and Haruka Takahashi }
\date{\today}
\address{Yokote Seiryo Gakuin High School, \\
147-1 Maeda, Osawa, Yokote, Japan
\ead{$*$ ztaro21@gmail.com}}

%\maketitle
\begin{abstract}
We present a sensitive diffusion cloud chamber which does not 
require any radioactive sources.  A major difference from a commonly used 
chamber is use of a heat sink as its bottom plate.  A result of 
a performance test of the chamber is given. 
\end{abstract}

\section*{Introduction}

In most countries, learning microscopic origin of nature is considered 
to be an important topic in science education.  However, there are not so many 
student experiments or demonstrations which exhibit the existence of
individual atom in direct and intuitive manner.  
The cloud chamber experiment is a rare exception; it provides most
direct and intuitive way to convince students existence of microscopic
particles.  Beautiful tracks of particles draw 
audience's attention and interest. Various chambers have been developed 
and used in classroom \cite{chambers}.  
Most of simple chambers seem to be based on the
results of Needles and Nielsen \cite{Needels}
and Cowan \cite{Cowan}, which use
a block of dry ice and a beaker filled with ethanol vapour.  
The great simplicity of their chamber enables students to make DIY
chamber at home or classroom.  Such a chamber works well if one uses a 
radioactive source.

\begin{figure}[htbp]
\begin{center}
 \includegraphics[width=0.7\textwidth]{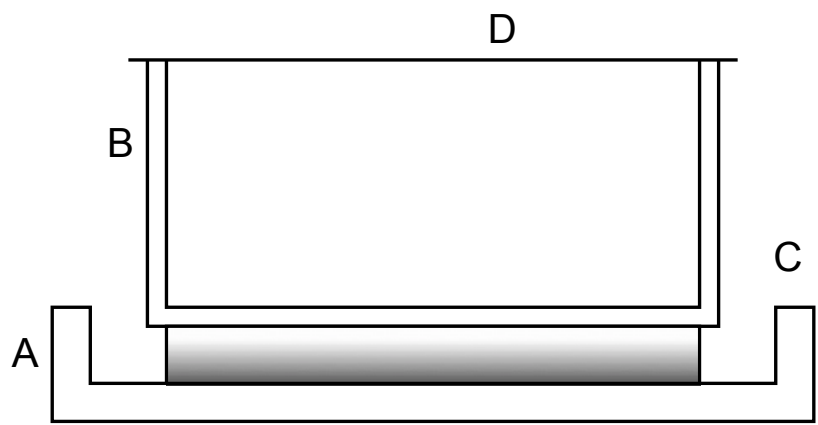}
\end{center}
\caption{A typical DIY chamber. A:styrofoam tray, B:glass container
 C:dry ice D:glass plate}
\label{typicalchamber}
\end{figure}

From spring 2008, one of the authors (S.Z.)
has been working on classroom experiment program
using cloud chamber. An one-day
experiment course had been given for local junior high school 
students.  From spring 2011, some students (rest of the authors) have
joined this project as a part of Super Science High School (SSH) activity. 
The typical setting of a chamber is shown in figure \ref{typicalchamber}.  
A chamber is not so large; smallest example is a glass laboratory dish
with radius 5cm and height 1cm.  A standard chamber we have used 
is a round glass container with radius 8.5cm and 8.0cm high \cite{rado}.
Such chambers require a \textit{radioactive source. }
It is very hard to find particle tracks of background radiation in this kind 
of chamber.  One will find few tracks in a minute only in very dark
room.

In Japan, people's attention to radioactivity is increased much after
the Fukushima Daiichi nuclear disaster following the Tohoku earthquake
and tsunami on 11 March 2011.  Large part of public unease or fear
about radioactivity is due to insufficient education about 
radioactivity, since curricula of primary and secondary schools 
lack serious and extensive study of radioactivity.
After the Tohoku earthquake, people have been nervous about
radioactivity.  In particular, parents do not wish their children
to join scientific activity using radioactive source, even if 
it is relatively safe one such as a gas mantle with thorium. 
Therefore, we think it is very important to develop a sensitive 
cloud chamber which {\it does not require any radioactive source.}
The chamber should be simple, small, portable and cheap since our aim is  
educational applications such as student experiment course or a workshop for
citizens in Tohoku area.

In this article, we would like to present a construction of a cloud
chamber works well without radioactive sources.
There is only minor modification from a typical chamber shown 
in this section, but its performance is remarkable.  We also 
present a result of a performance test.

\section*{The construction of a cloud chamber}
\begin{figure}[htbp]
\begin{center}
\includegraphics[width=0.7\textwidth]{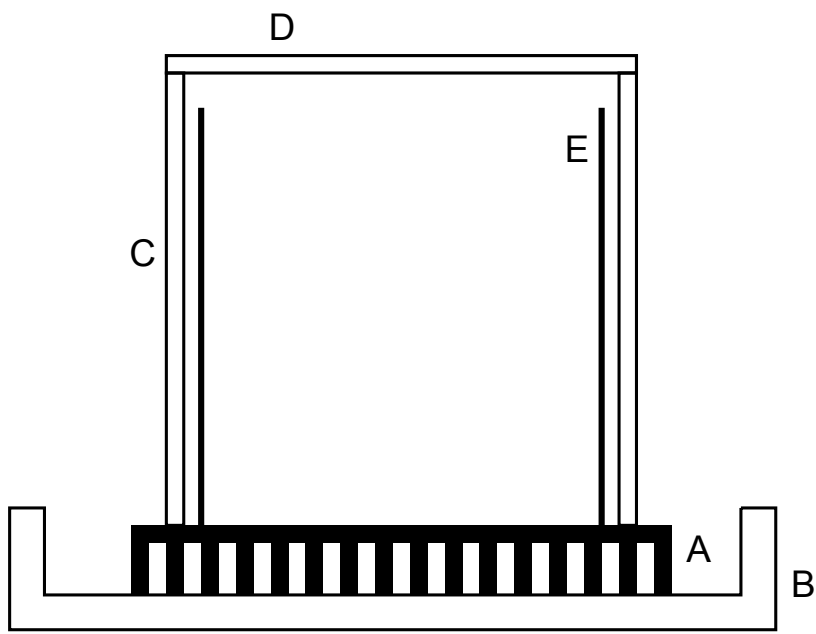}
\caption{Construction of a cloud chamber. A:heat sink,  
B:styrofoam tray, C:acrylic container, D:acrylic plate, E:black felt. }
\label{fig:chamber} 
\end{center}
\end{figure}
The construction of our cloud chamber is shown in figure \ref{fig:chamber}.
A major difference from a common chamber is use of a black anodised aluminium
heat sink as a bottom plate instead of a metal plate. 
Its dimension is 134mm $\times$ 134mm $\times$ 20mm, weight is 
%622 kg, 
622 g, and a side with fins is placed downward.
As we see later, the heat sink greatly improves performance of our
chamber.   Our method has an additional advantage that the heat sink 
plays a role of supporting base of the chamber which prevents direct
contact of side wall with refrigerant. 
Clear walls without a bottom and a top are placed on the heat sink. 
Walls and top are made of square, 5mm thick acrylic plates. Top plate is
removable.  A piece of black felt E is placed along 
three sides of a box.  Upper half of it is soaked with ethanol.  
Total amount of ethanol is about 10g for each experiment.
Light from PC projector is introduced from a face without a 
piece of black felt.   All of the above instruments are placed on a
shallow styrofoam tray B.  Liquid nitrogen can be directory poured into
this box to this tray.  Powder of dry ice can be used also.  In
this case, a heat sink is placed above powdered dry ice.  
\begin{figure}[htbp]
\begin{center}
\includegraphics[width=0.7\textwidth]{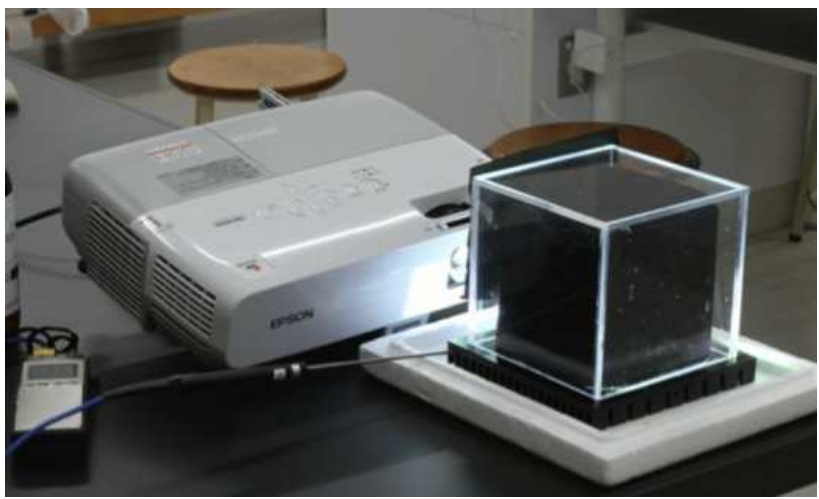}
\caption{A cloud chamber at work.}
\label{fig:working} 
\end{center}
\end{figure}
In the placement mentioned above, tracks of particles can be observed from
above. Figure \ref{fig:working} is a photo of our cloud chamber in
operation. A PC projector is slanted to obtain fine view of particle
tracks.  

%%% camera description is moved to next section

In closing this section, we would like to mention that our result is not
new. An advantage of our chamber is good performance in spite of its smallness and simpleness. 
In fact, various examples of sensitive chambers have been known in Japan.  
Very large and expensive chambers for display are 
available in some science museums \cite{rado}.  A glass chamber
presented in the beginning of this section is also available 
as a product by Rado Ltd.\  \cite{rado}.  A use of a small heat sink with
liquid nitrogen is first presented in \cite{Mori}.  Another example of
a sensitive chamber is given in \cite{Hayashi}.   

\section*{Operation}

Surprisingly, the very simple construction presented in the previous
section is enough to see many particle tracks without any radioactive 
sources.  External electric field is also unnecessary.
We would like to describe operation of a cloud
chamber.  To begin with, liquid nitrogen was
poured to the styrofoam box.  Our lab was at room
temperature, so the temperature of the bottom heat sink rises as liquid
nitrogen evaporates. During the observation, this change of the temperature 
was measured by a thermocouple.
\begin{figure}[htbp]
\begin{center}
  \includegraphics[width=.7\textwidth]{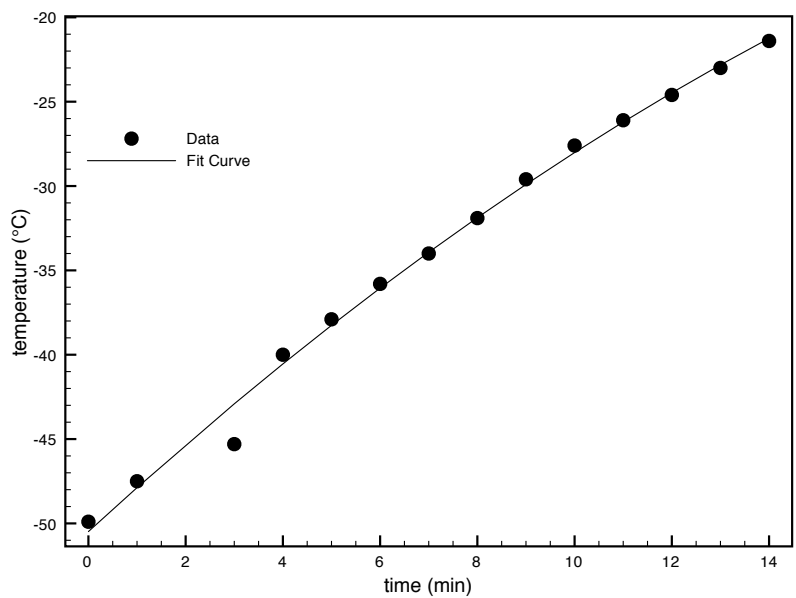}
\caption{Change of temperature of the bottom heat sink}
\label{fig:bottomtemp}
\end{center}
\end{figure}
A result is shown in figure \ref{fig:bottomtemp}. Obtained data fits well to 
a quadratic curve.
%\begin{equation}
% T = -0.040 t^2 +2.6 t - 51,
%\end{equation}
%where $T$ and $t$ are temperature (degrees Celsius) and 
%elapsed time $t$ (min.) respectively.

\begin{figure}[htbp]
\begin{center}
  \includegraphics[width=\textwidth]{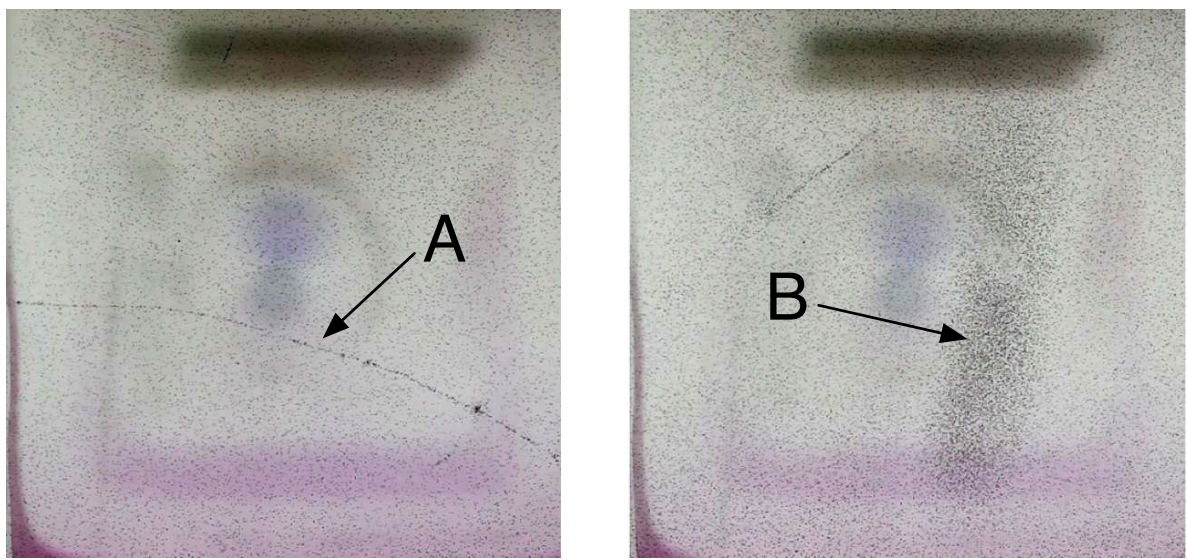}
\caption{Particle tracks seen in the chamber.  A: a thin, long track, B:
an alpha particle track.  Black and white are converted.}
\label{fig:alphabeta}
\end{center}
\end{figure}

Particle tracks begin to appear in few minutes.
Here we present our result with a video \cite{video} uploaded to
YouTube.  Figure \ref{fig:alphabeta} is captured image from the video.
Most thin and wiggy tracks, such as the left picture of 
figure \ref{fig:alphabeta} frequently seen in the chamber, can be
considered as those of beta particles.
 We can see long tracks since the bottom plate is cooled
uniformly.  Some of long tracks are slightly
curved even though no external magnetic field is imposed.  We 
think it is due to multiple collision with atoms in the air.
An alpha particle also can be observed as a short, thick line 
such as the right picture in figure \ref{fig:alphabeta}. 
One can see that many droplets are formed
along a track and fall down toward bottom of the chamber.  

Although many tracks are seen in the video, there should be 
no influence of Fukushima Daiichi nuclear disaster, 
since in our city (Yokote, Akita, Japan) has never 
observed apparent rise of air dose rate since 11th March
\footnote{If we do similar experiment in Fukushima where air dose rate is
still high, more tracks will be observed. 
A video \cite{Fukushima} uses soil brought from three places
near the Fukushima Daiich power plant.}.

\begin{figure}[htbp]
\begin{center}
  \includegraphics[width=.7\textwidth]{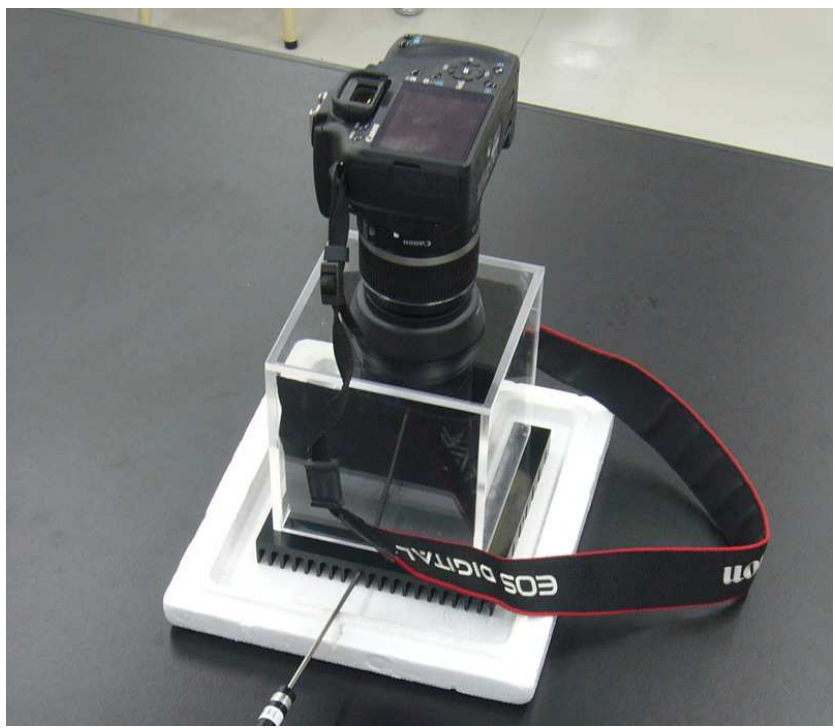}
\caption{A digital SLR camera mounted on the top of the 
cloud chamber. }
\label{fig:camera}
\end{center}
\end{figure}

Obtaining of high quality video is important both for scientific
analysis and publication on the web.  The later is especially
important in for education, since lower quality videos found in 
the web are not enough to tell the beauty and excitement that people
must feel in this experiment. We find 
that a digital SLR camera (in our case, EOS kiss X3) is very 
useful for this purpose. It has larger CCD and better lens required for 
high quality video.  In our experiment, a camera is directly put on the top plate as in 
figure \ref{fig:camera}, but use of rigid is better if available.

\section*{Performance evaluation}

\begin{figure}[htbp]
\begin{center}
  \includegraphics[width=.7\textwidth]{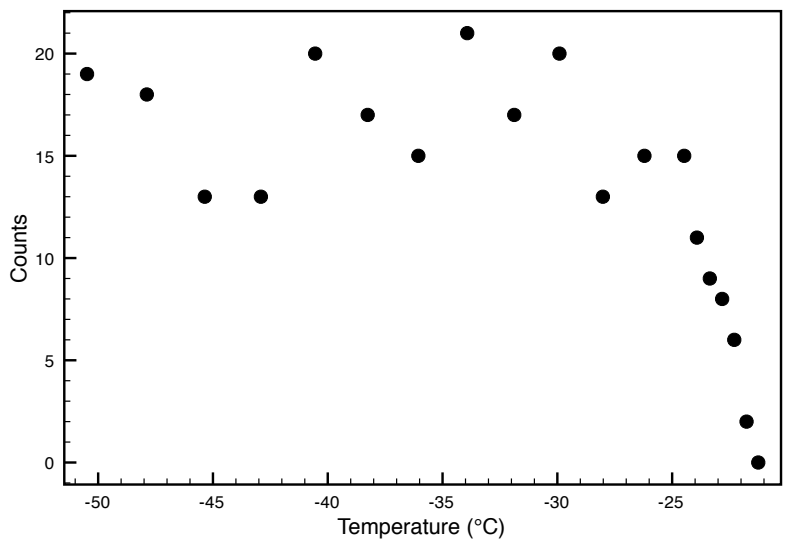}
\end{center}
\caption{A number of particle tracks per 10 seconds}
\label{fig:countpermin}
\end{figure}

A qualitative evaluation of the performance of a chamber will be 
useful for development of new cloud chamber and finding better 
working condition.  In fact, some test runs tells us that
numbers of particle tracks seems to differ for each run.  Using
a recorded video \cite{video}, we 
counted all tracks seen in the chamber regardless of its length,
width and shape. Therefore alpha particles, beta particles and other
particles are not distinguished. 
%From the video, one can see that beta
%particles are dominant. 
Figure \ref{fig:countpermin} shows a
relation between count per 10 seconds and elapsed time.  
We observe that, although fluctuation is large, the count is independent of
the bottom temperature until it reaches to `critical'
value.  Our observation shows the fall of the count begins around $-25$
degrees. 
\begin{figure}[htbp]
\begin{center}
\includegraphics[width=0.9\textwidth]{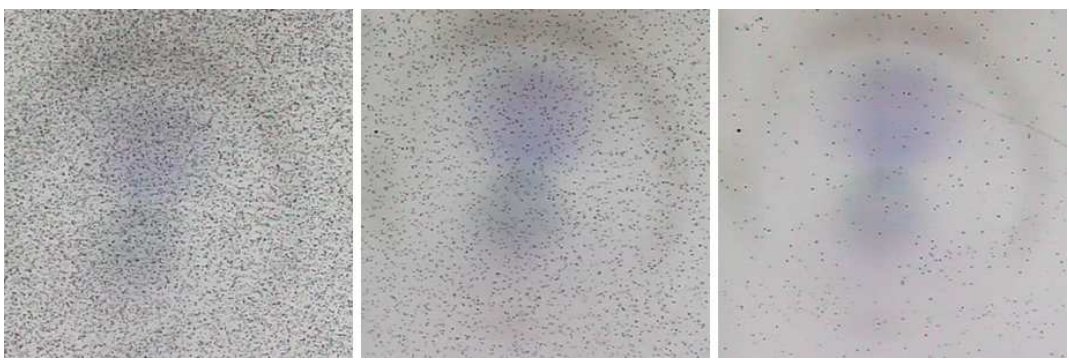}
\caption{Photos of condensed droplets in the chamber.  From the left, 
the bottom temperatures are
 $-51$, $-35$  and $-26$ degrees respectively.  
Black and white are converted. }
\label{fig:droplets} 
\end{center}
\end{figure}
While a number of particle tracks does not change below critical
temperature, a number of background 
droplets affects visibility of particle tracks.
Many droplets are seen at lower temperature as shown in the left
of figure \ref{fig:droplets}.  Contrast of whole image is reduced at such 
low temperature.  We also found that alpha particle tracks become
thinner at lower temperature, as mentioned in \cite{Slatis}.
On the other hand, it becomes hard to identify particle tracks at
high temperature.  Therefore, the best temperature range for observation
should be middle region in figure \ref{fig:countpermin}.

\section*{Perspective}

We expect our cloud chamber will be useful for various application to education,
for example, a student research project or demonstrations in public.
In particular, it will be important to measure numbers of particle tracks 
in a city  where  dose rate is still high.  Also not examined in this paper,
the high quality video enables us various computer analysis such as 
qualitative evaluation of amount of condensation or automatic counting of
particle tracks.  Such kind of analysis will be useful for a 
student's research projects.

\section*{Acknowledgments}
We would like to thank Hiroki Kanda of Tohoku university
for useful comments at the beginning
of our project, Ichiro Itoh of Yokote Seiryo Gakuin H.S. for 
providing a voltage multiplier, Ryoko Yamaishi for support
in lab.  
%% added 17 Feb. 
We also thank Kenichiro Aoki of Keio university for careful 
reading of the manuscript. 
This work is supported by
Saito Kenzo honour fund and the Super Science High School 
funding from Japan science and technology agency.

\end{document}